\documentclass[apj]{emulateapj}             %%%%%%%%%%%%%%%%%%% 2columns, comment for ApJ submission
\usepackage{graphicx}                        %%%%%%%%%%%%%%%%%%% comment for ApJ submission%
\usepackage{txfonts}                        %%%%%%%%%%%%%%%%%%% comment for ApJ submission
\usepackage{natbib}                        %%%%%%%%%%%%%%%%%%% comment for ApJ submission
\usepackage{longtable}                        %%%%%%%%%%%%%%%%%%% comment for ApJ submission
\def \inte {{$INTEGRAL$}}
\def \xmm {{\em XMM--Newton}}

\def \sw {{\em Swift}}
\def \src {\mbox{IGR~J11215--5952}}

\def \hcm {\hbox {\ifmmode $ atom cm$^{-2}\else atom cm$^{-2}$\fi}}

\def \ATel {The Astronomer's Telegram}

\slugcomment{To appear in ApJ}
\shorttitle{System geometry in IGR J11215$-$5952}
\shortauthors{Romano et al.}

\begin{document}
\title{Disentangling the system geometry  of the Supergiant Fast 
X-ray Transient IGR J11215$-$5952 with {\em Swift} }

\author{P.~Romano\altaffilmark{1}, 
L.~Sidoli\altaffilmark{2}, 
G.~Cusumano\altaffilmark{1},  
S.~Vercellone\altaffilmark{2,1},  
V.~Mangano\altaffilmark{1},  
H.~A.~ Krimm\altaffilmark{3,4} 
}
  \altaffiltext{1}{INAF, Istituto di Astrofisica Spaziale e Fisica Cosmica, 
	Via U.\ La Malfa 153, I-90146 Palermo, Italy} 
 \altaffiltext{2}{INAF, Istituto di Astrofisica Spaziale e Fisica Cosmica, 
	Via E.\ Bassini 15,   I-20133 Milano,  Italy}
   \altaffiltext{3}{CRESST/Goddard Space Flight Center, Greenbelt, MD, USA}
 \altaffiltext{4}{Universities Space Research Association, Columbia, MD, USA}

\begin{abstract}

IGR J11215-5952 is a hard X--ray transient source discovered in 2005 April with 
$INTEGRAL$ and a member of the new class of High Mass X--ray Binaries, 
the Supergiant Fast X--ray Transients (SFXTs). 
While $INTEGRAL$ and $RXTE$ observations have shown that the outbursts occur
with a periodicity of $\sim330$ days, \sw\  data have recently demonstrated that 
the true outburst period is $\sim 165$ days. 
\src\ is the first discovered SFXT displaying periodic outbursts, 
which are possibly related to the orbital period.
The physical mechanism responsible for the X-ray outbursts in SFXTs is still debated. 
The main hypotheses  proposed to date involve the structure of the companion wind
or gated mechanisms related to the properties of the compact object.
We test our proposed model which explains  the outbursts from SFXTs as being due to
the passage of the neutron star inside the equatorially enhanced wind from the supergiant companion. 
We performed a Guest Investigator observation with \sw\ 
that lasted 20\,ks and several follow-up Target of Opportunity (ToO) 
observations, for a total of $\sim32$\,ks, during the expected ``apastron'' 
passage (defined assuming an orbital period of $\sim$330 days), 
between  2008 June 16 and  July 4. 
The characteristics of this ``apastron'' outburst are quite similar to those 
previously observed during the ``periastron'' outburst of 2007 February 9.
The mean spectrum of the bright peaks can be fit with an absorbed power law model with 
a photon index of 1 and an absorbing column of $\sim 10^{22}$ cm$^{-2}$. 
This outburst reached luminosities of $\sim 10^{36}$ erg\,s$^{-1}$ (1--10\,keV),
comparable with the ones measured in 2007. The light curve can
be modelled with the parameters obtained by Sidoli et al.\ (2007) for the 
2007 February 9 outburst, although some differences can be observed in its shape. 
The properties of the rise to this new outburst and the comparison with the previous outbursts
allow us to suggest that the true orbital period of \src\ is very likely 164.6~days, and that the orbit
is eccentric, with the different outbursts produced at the periastron passage,
when the neutron star crosses the inclined equatorial wind from the supergiant companion. 
Based on a ToO observation performed on 2008 March 25--27, we can exclude that 
the period is 165/2 days. 
\end{abstract}


%%%%%%%%%%%%%%%%%%%%%%%%%%%%%%%%%%%%%%%%%%%%%%%%%%%%%%%%%%%%%%%%%%%%%
\section{Introduction}
%%%%%%%%%%%%%%%%%%%%%%%%%%%%%%%%%%%%%%%%%%%%%%%%%%%%%%%%%%%%%%%%%%%%

	\begin{figure*}%%%%%%%%%%%%%%%%%%%%%%%%%%%%%%%%%%%%%%%%%%%%% FIGURE 1
	 	\includegraphics[width=18cm,height=17cm]{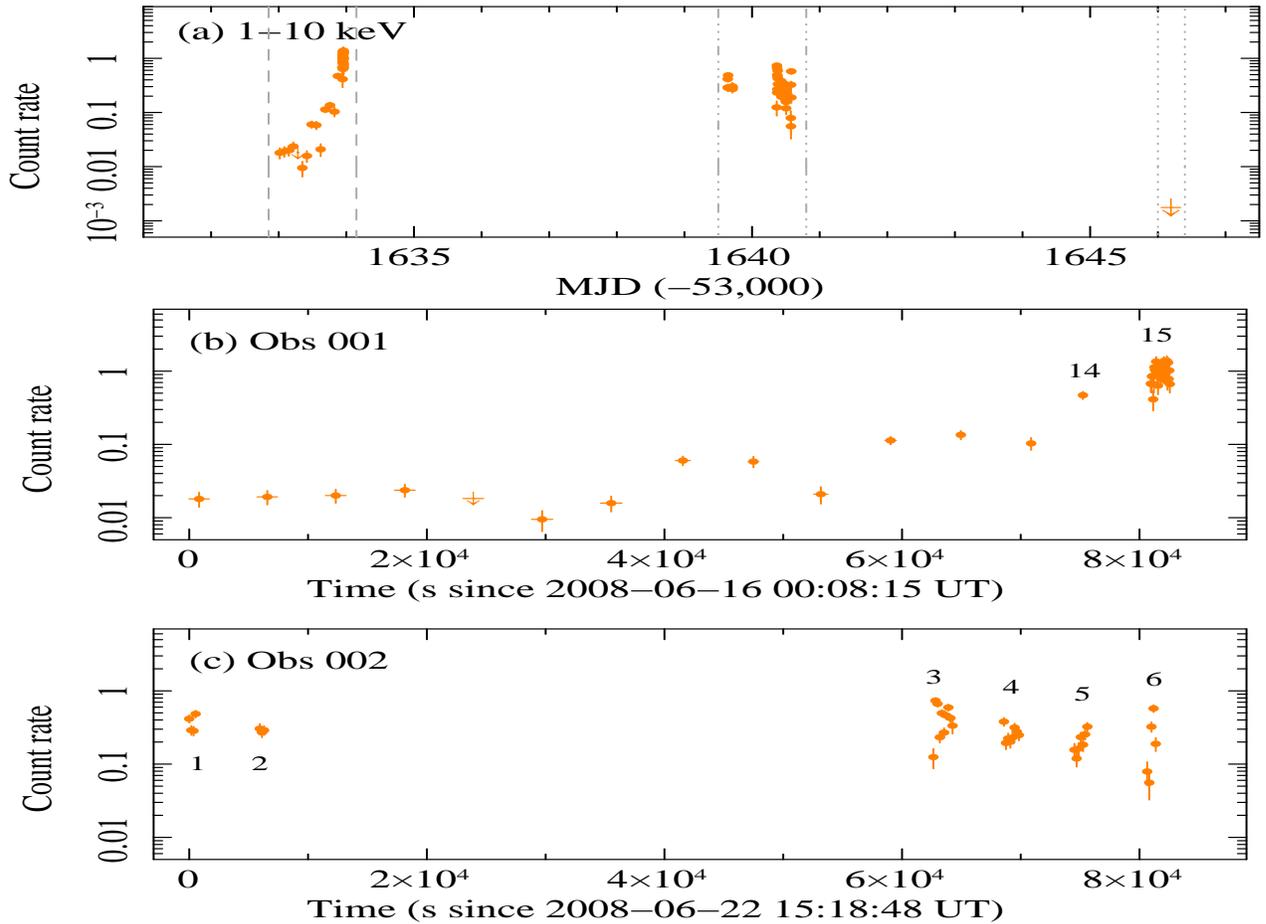}
	\vspace{-3cm}
		\caption{XRT light curves, background-subtracted and corrected 
		for pile-up, PSF losses, and vignetting  (starting on 2008-06-14 00:00:00 UT). 
		{\bf a)} 1--10\,keV light curve for the whole campaign.
		Filled circles are detections (S/N$>$3), 
		while downward-pointing arrows are 3-$\sigma$ upper limits. 
		The vertical lines mark the boundaries of the three observations. 
		{\bf b)} Detail of observation 1 (referred to MJD 54633.006, 2008-06-16 00:08:15 UT). 
		The numbers mark the positions of the orbits on which we performed spectroscopic analysis. 
		(Sect.~\ref{igr112p3:anal_apastron}).
		{\bf c)} Detail of observation 2 (referred to MJD 54639.638, 2008-06-22 15:18:48 UT). 
		The numbers mark the positions of the orbits. 
                Note that where no data are plotted, no data were collected. 
		}
 		\label{igr112p3:fig:lcv}
	\end{figure*}

\object{IGR~J11215--5952} \citep{Lubinski2005} is an accreting pulsar 
($P_{\rm spin}=186.78\pm0.3$\,s, \citealt{Swank2007}) and a 
member of the new class of High Mass X-ray Binaries (HMXBs) of
the Supergiant Fast X--ray Transients (SFXTs),  which we define as 
transient and flaring X--ray sources with a firm association (via optical spectroscopy) 
with an O or B supergiant. In particular, this X-ray transient has a B1 
supergiant companion \citep{Negueruela2005b,Masetti2006,Steeghs2006}
located at a distance of about 6 kpc. 

The analysis of archival \inte\ observations of the source field 
led to the discovery of a recurrence period of $\sim$330~days in the X--ray activity 
\citep*[hereafter Paper I]{SidoliPM2006}, probably
linked to the orbital period of the binary system, with the outbursts
triggered near the periastron passage.
This periodicity was later confirmed with $RXTE$/PCA  in 2006, which
observed a new outburst 329~days after the previous one \citep{Smith2006a}.
The X--ray spectrum (5--100\,keV) observed with \inte\ was well fitted by a hard power-law
with a high energy cut-off around 15~keV (Paper~I), with peak 
luminosity of $\sim 3 \times$10$^{36}$~erg~s$^{-1}$ (Paper~I).

Exploiting the unique predictability of the outbursts, we performed a 
target of opportunity (ToO) observation with \sw{} with the main aim of 
monitoring the source behavior around the time of the fifth  
outburst, expected on 2007 February 9.
This led to the most complete observation of a SFXT outburst 
\citep[][hereafter Paper~II]{Romano2007}, 
that lasted 23 days for a total on-source exposure of $\sim73$\,ks. 
These \sw{} observations constitute a unique data-set, 
thanks to the combination of sensitivity and time coverage, and they allowed a 
study of \src\ from outburst onset to almost quiescence.  
We found that the accretion phase lasts longer than previously thought on the basis 
of lower-sensitivity instruments observing only the brightest flares, and that short
outbursts (the flares lasting minutes to few hours) are actually part of a
much longer outburst event (lasting several days). 
We also found that the spectrum during the brightest flares is described well by an
absorbed power law with a photon index of 1 and $N_{\rm H} \sim 1 \times 10^{22}$ cm$^{-2}$  
and derived a 1--10\,keV peak luminosity of $\sim $10$^{36}$~erg~s$^{-1}$. 

A second monitoring ToO  with \sw/XRT, performed around the supposed 
apastron passage (predicted on 2007 July 24), led to the discovery of a new 
unexpected outburst, after $\sim 165$ days from the latest outburst  
(\citealt{Romano2007apastron};  \citealt[][hereafter Paper~III]{Sidoli2007}). 
In Paper III we discussed the different implications of these findings
and the possible geometries for this binary system and for the SFXTs in
general.

Here we report the results of a 20\,ks observation with 
\sw, planned to cover the onset of the new outburst (an ``apastron'' outburst), 
329 days from the 2007 July 24 outburst. 
We also report on two Target of Opportunity Observations  with \sw/XRT
at a quarter of the period reported in Paper~I (defined assuming an orbital 
period of $P=329$ days).

  %%%%%%%%%%%%%%%%%%%%%%%%%%%%%%%%%%%%%%%%%%%%%%%%%%%%%%%%%%%%%%%%%%%%
  \section{Observations and Data Reduction\label{igr112p3:dataredu}}
  %%%%%%%%%%%%%%%%%%%%%%%%%%%%%%%%%%%%%%%%%%%%%%%%%%%%%%%%%%%%%%%%%%%%

	\begin{figure}[t]%%%%%%%%%%%%%%%%%%%%%%%%%%%%%%%%%%%%%%%%%%%%% FIGURE 2
	\vspace{-2.0cm}
	 	\includegraphics[angle=0,width=9cm]{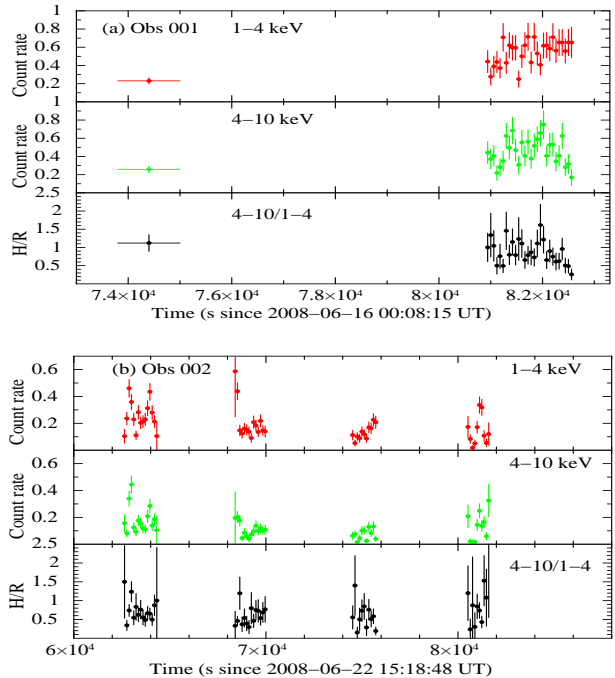}
	\vspace{-1.5cm}
 		\caption{ {\bf (a)} Detail of the last two orbits of observation 001 
                        ($>60$\,s bins), 
			showing the 1--4\,keV, 4--10\,keV count rates 
			(top and middle panel) and the hardness ratio 4--10/1--4 (bottom).
		        {\bf (b)} Same as a), for the last 4 orbits of observation 002 
			(120\,s bins). 
		        }
                \label{igr112p3:fig:hr}
	\end{figure}

Table~\ref{igr112p3:tab:alldata} reports the log of the \sw/XRT observations 
used for this work. 
% 
%%%%%%%% 
The first set of observations were obtained as a ToO to investigate the
length of the outburst period, at a quarter of the period reported in Paper I ($P=329$ days). 
The first observation (00030881033) lasted 1.8\,ks, starting on 2008-03-25 at 00:16:42 UT, 
the second (00030881034) lasted 1.9\,ks, starting on 2008-03-27 at 00:26:47 UT,
for a total of 3.8\,ks. 
%
%
%%%%%%%% 
The second set of observations were obtained as part of the Guest Investigation 
program, (00090005001, hereafter 001, 20\,ks) and as follow-up ToOs 
(00090005002 and 00090005003, hereafter 002 and 003, and 00030881035), 
for a total of 32\,ks. 

The XRT data were processed with standard procedures 
({\tt xrtpipeline} v0.11.6), filtering and screening criteria by using 
FTOOLS in the {\tt Heasoft} package (v.6.4).  
Given the low rate of the source during the whole campaign,  
we only considered photon counting (PC) data and further selected 
XRT event grades 0--12 (\citealt{Burrows2005:XRT}). 
Pile-up correction was necessary in observation 001, hence
we adopted an annular source extraction 
region with radii 3 and 30 pixels. 
Since the source Cen X-3, located 45.7 arcmin from \src\ produced 
single reflection rings in the XRT images, the background was chosen as 
a circular region at the same radial distance as \src\ from the center of 
the rings, and with a 60-pixel radius. 
The source was not detected in the third segment (003), with a 3-$\sigma$ upper 
limit on the unabsorbed flux (for $\Gamma=1.0$, $N_{\rm H} = 10^{-22}$ cm$^{-2}$, 
see Paper~II) of $5.0\times 10^{-13}$  erg cm$^{-2}$ s$^{-1}$, or 
$2.3\times10^{33}$ erg s$^{-1}$ (at 6.2\,kpc, 1--10 keV). 
Nor was it detected in observation 00030881035 (3-$\sigma$ upper limit at 
$4.7\times 10^{-12}$  erg cm$^{-2}$ s$^{-1}$, or $2.1\times10^{34}$ erg s$^{-1}$).
The remainder of the in-depth analysis therefore only refers to observations 
001 and 002. 
Ancillary response files were generated with {\tt xrtmkarf},
and they account for different extraction regions, vignetting, and
PSF corrections. We used the latest spectral redistribution matrices
(v010) in CALDB. 
For timing analysis, the arrival times of XRT events were
converted to the Solar System barycentre with the task 
{\tt barycorr} and source events were extracted from a 
circular region (20 pixels radius).

The BAT always observed \src\ simultaneously with XRT, so 
survey data products, in the form of Detector Plane
Histograms (DPH) with typical integration time of 
$\sim 300$\,s, are available. Furthermore, event data 
were captured as part of the GI proposal requests up to the
maximum of $\sim 20$\, minutes per orbit (15 orbits total) 
allowed by the large telemetry bandwidth required. 
The BAT data were analyzed using the standard BAT analysis 
software distributed within FTOOLS.
We never detected the source above a
signal-to-noise ratio (S/N) threshold of 5, and we obtained 
5-$\sigma$ flux upper limits of 
$8.17\times10^{-11}$ erg cm$^{-2}$ s$^{-1}$ (15--25\,keV),   
$5.95\times10^{-11}$ erg cm$^{-2}$ s$^{-1}$ (15--30\,keV), and  
$5.34\times10^{-11}$ erg cm$^{-2}$ s$^{-1}$ (15--50\,keV). 
These were obtained with a comparison with a Crab on axis observation. 
This is consistent with the extrapolation at the high energies
of the XRT data fit with an absorbed power-law with a high energy cutoff
at 15\,keV (and $\Gamma=0.5$) drawn from the {\em INTEGRAL}/ISGRI 2003 July data fit
(Paper I).

All quoted uncertainties are given at 90\% confidence level for 
one interesting parameter unless otherwise stated. 
The spectral indices are parameterized as  
$F_{\nu} \propto \nu^{-\alpha}$, 
where $F_{\nu}$ (erg cm$^{-2}$ s$^{-1}$ Hz$^{-1}$) is the 
flux density as a function of frequency $\nu$; 
we also use $\Gamma = \alpha +1$ as the photon index, 
$N(E) \propto E^{-\Gamma}$ (ph cm$^{-2}$ s$^{-1}$ keV$^{-1}$).

  %%%%%%%%%%%%%%%%%%%%%%%%%%%%%%%%%%%%%%%%%%%%%%%%%%%%%
  \section{Analysis and Results\label{igr112p3:dataanal}}
  %%%%%%%%%%%%%%%%%%%%%%%%%%%%%%%%%%%%%%%%%%%%%%%%%%%%%

  %%%%%%%%%%%%%%%%%%%%%%%%%%%%%%%%%%%%%%%%%%%%%%%%%%%%%
  \subsection{The 2008 March observations ($P/4$)\label{igr112p3:anal_p4}}
  %%%%%%%%%%%%%%%%%%%%%%%%%%%%%%%%%%%%%%%%%%%%%%%%%%%%%

The source was not detected in either observation. 
The 3-$\sigma$ upper limit on the unabsorbed flux of the cumulative observation 
(obtained assuming the spectral parameters based on the 2007 February outburst reported in 
Paper~II; $\Gamma=1.0$, $N_{\rm H} = 10^{-22}$ cm$^{-2}$) 
is $9.1\times 10^{-13}$ erg cm$^{-2}$ s$^{-1}$  (1--10 keV), 
which translates into a 1--10 keV luminosity of $4.1\times10^{33}$ erg s$^{-1}$  at 6.2\,kpc. 
The 3-$\sigma$ upper limit on the unabsorbed flux can be compared with the detections at 
$2.3\times 10^{-12}$, $9.2\times 10^{-12}$, and $9.5\times 10^{-12}$ erg cm$^{-2}$ s$^{-1}$  
obtained during the 2007 February outburst at $T-48$ hours, $T-24$ hours $T+24$ hours, 
respectively.

  %%%%%%%%%%%%%%%%%%%%%%%%%%%%%%%%%%%%%%%%%%%%%%%%%%%%%
  \subsection{The 2008 June observations (``apastron'')\label{igr112p3:anal_apastron}}
  %%%%%%%%%%%%%%%%%%%%%%%%%%%%%%%%%%%%%%%%%%%%%%%%%%%%%

As done in Paper~II, we extracted light curves in 
the 1--10\,keV, 1--4\,keV,  and 4--10\,keV bands, 
but discarded the 0.2--1\,keV band because of its 
significantly lower signal with respect to the other bands.  
The light curves were corrected for Point-Spread Function 
(PSF) losses, due to the extraction region geometry, 
bad/hot pixels and columns falling within this region, 
and for vignetting, by using
the task {\tt xrtlccorr}, which generates 
an orbit-by-orbit correction based on the instrument map. 
We then subtracted the scaled background rate in each band from their 
respective source light curves and calculated the 4--10/1--4 hardness ratio. 

Figure~\ref{igr112p3:fig:lcv}a shows the complete light curve of \src\ 
throughout our GI program and follow-up ToOs in MJD units. 
Panels (b) and (c) show a detail of observations 001 and 002, respectively,
in seconds since the start of the observation. 
Panel (b), in particular, shows the rise to the outburst, with 
an increase in count rate by a factor of $\sim 32$ in less than 8 hours
(orbit 10 with respect to orbit 15), by a factor of $\sim 2.5$ in 1.7 hours 
(orbit 14 with respect to 15).
However, no significant variation in the hardness ratio (Fig.~\ref{igr112p3:fig:hr}) 
can be evidenced. 
Indeed, fitting the hardness ratio as a function of time to a constant model
we obtain a value of $0.70\pm0.08$ ($\chi^2_{\nu}=1.1$ for 28 degrees of freedom, d.o.f.) 
for orbit 14 and 15 of observation 001 (Fig.~\ref{igr112p3:fig:hr}a), and 
a value of $0.51\pm0.05$ ($\chi^2_{\nu}=1.08$ for 49 d.o.f.) for
orbits 3--6 of observation 002 (Fig.~\ref{igr112p3:fig:hr}b).

\begin{figure}[t]%%%%%%%%%%%%%%%%%%%%%%%%%%%%%%%%%%%%%%%%%%%%% FIGURE 3
	 	\includegraphics[angle=270,width=9.0cm]{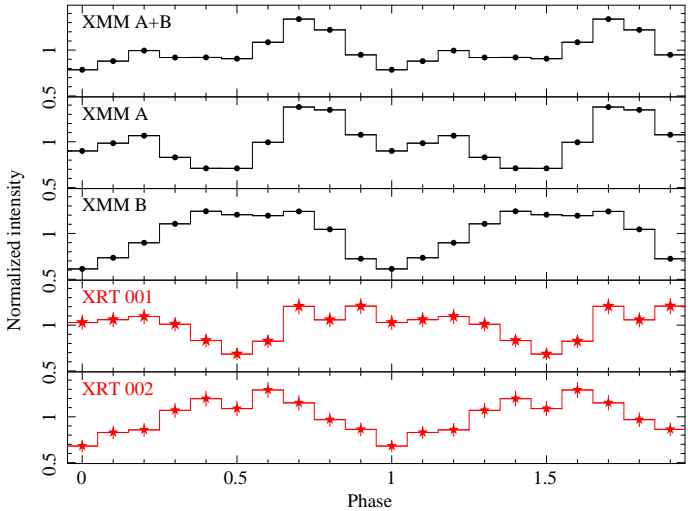}
 		\caption{Folded 0.2--10\,keV light curves of observation 001 and 002 (red
		stars, bottom two panels) using a period of 186.78\,s. 
		The marks A and B refer to the convention adopted for the \xmm\ data (black
		filled circles)
		obtained during the 2007 February observation in \citet{Sidoli2007},
		(A bright state, B faint state, A$+$B whole \xmm\ observation).  
 		}
                \label{igr112p3:fig:foldlcv}
	\end{figure}

We sought for the \src\ spin periodicity by using epoch 
folding techniques, on the combined 001 and 002 observations, 
finding a period of $P_{\rm spin}=186.8\pm0.1$\,s, which is consistent with that 
derived from $RXTE$/PCA data \citep[][$186.78\pm0.3$\,s]{Swank2007}
and by \xmm\ data in Paper~III ($186.94\pm0.58$\,s,). 
We folded the data of observation 001 and 002 (separately) at the 
period of 186.78\,s  and obtained the 0.2--10\,keV 
light curves shown in Fig.~\ref{igr112p3:fig:foldlcv}. For ease of comparison, 
we also report the folded light curve in the bright (A) and faint (B) parts 
of the \xmm\ light curves reported in Paper~III. 
We note that the early (and brighter) part of the XRT 001 observation 
shows a folded profile in phase with the bright part (A) of the  
\xmm\ light curve, while the 002 observation is in phase with the faint (B) part.

	\begin{figure}[t]%%%%%%%%%%%%%%%%%%%%%%%%%%%%%%%%%%%%%%%%%%%%% FIGURE 4
 	\vspace{-1.5cm}
	 	\includegraphics[angle=0,width=9cm]{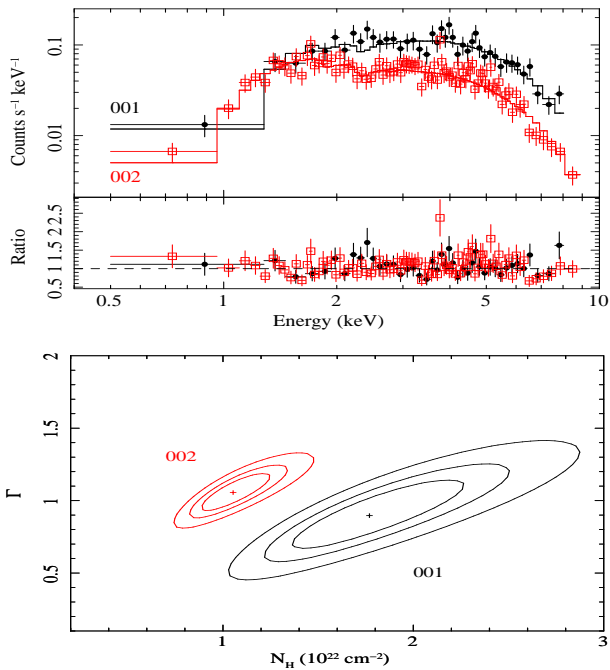}
 	\vspace{-1.5cm}
		\caption{XRT time-selected spectroscopy of the 2008 June outburst. 
		{\bf Upper panel:} Top: observation 001, orbit 15 spectrum 
		(2008 June 16, black filled circles), 
		and observation 002 (2008 June 22, red empty squares). 
		 Bottom: data/model ratio. 
		{\bf Lower panel:} $\Delta \chi^2 =2.3, 4.61, 9.21$ contour levels for the 
		column density in units of $10^{22}$ cm$^{-2}$  
		vs.\ the photon index, with best-fit values indicated by crosses
		for observation 001 (orbit 15) and observation 002. 
		}
                \label{igr112p3:fig:spec}
	\end{figure}

Upon examination of the light curve presented in Fig.~\ref{igr112p3:fig:lcv} and the 
available counting statistics, we selected different time bins over which we accumulated 
spectra. 
These include:  
{\it i)} the low-level  phase before the 2008 June 16 outburst, when the mean count rate 
was $\sim 0.01$ counts s$^{-1}$  (orbits 1 through 13 of observation 001); 
{\it ii)} the intermediate phase (mean CR $\sim 0.5$ counts s$^{-1}$, orbit 14 of observation 001); 
{\it iii)} the beginning of the outburst (orbit 15 of observation 001);
{\it iv)} each of the orbits in observation 002;
{\it v)} the whole observation 002. 
The data were generally rebinned with a minimum of 
20 counts per energy bin to allow $\chi^2$ fitting. 
However, in several instances the Cash statistic \citep{Cash1979} and spectrally 
unbinned data were used, due to the low counting statistics.  
The spectra were fit with XSPEC (v11.3.2) in the 0.5--9\,keV energy range,
adopting an absorbed power law model. 

The best fit parameters are reported in Table~\ref{igr112p3:tab:specfits} 
along with the mean luminosity of each time selection. 
Fitting the spectrum of the first flare (orbit 15 of observation 001)  
we obtain a photon index 
$\Gamma=0.90_{-0.25}^{+0.27}$ and an absorbing column density of 
$N_{\rm H}=(1.77_{-0.44}^{+0.55})\times 10^{22}$ cm$^{-2}$ 
($\chi^2_{\rm red}=0.861/40$ d.o.f.), 
while observation 002 yielded 
$\Gamma=1.06_{-0.14}^{+0.14}$ and an absorbing column density of 
$N_{\rm H}=(1.05_{-0.18}^{+0.21})\times 10^{22}$ cm$^{-2}$.  
Fig.~\ref{igr112p3:fig:spec} shows the spectra and the 
contour levels for the column density vs.\ the photon index.

%%%%%%%%%%%%%%%%%%%%%%%%%%%%%%%%%%%%%%%%%%%%%%%%%%%%%%%%%%%%%%%%%
\section{Discussion and Conclusions\label{igr112p3:discussion}}
%%%%%%%%%%%%%%%%%%%%%%%%%%%%%%%%%%%%%%%%%%%%%%%%%%%%%%%%%%%%%%%%%

	\begin{figure*}[t]%%%%%%%%%%%%%%%%%%%%%%%%%%%%%%%%%%%%%%%%%%%%% FIGURE 5
	 	\includegraphics[angle=270,width=18cm]{f5.eps}
 		\caption{Complete XRT light curve, in the 1--10\,keV range,
		obtained from all \sw\ pointed observations. 
		Filled circles are detections (S/N$>$3), 
		triangles marginal detections ($2<$S/N$<3$, for the 2007 data), 
		while downward-pointing arrows are 3-$\sigma$ upper limits. 
		Grey data (before MJD 54400) have already been presented in 
                \citet{Romano2007} and \citet{Sidoli2007};  
		light grey indicates ``periastron'' outbursts, dark grey ``apastron''
		outbursts (defined assuming an orbital period of $P=329$ days). 
		}
                \label{igr112p3:fig:historical_lcv}
	\end{figure*}

	\begin{figure*}[t]%%%%%%%%%%%%%%%%%%%%%%%%%%%%%%%%%%%%%%%%%%%%% FIGURE 6
	 	\includegraphics[angle=270,width=18cm]{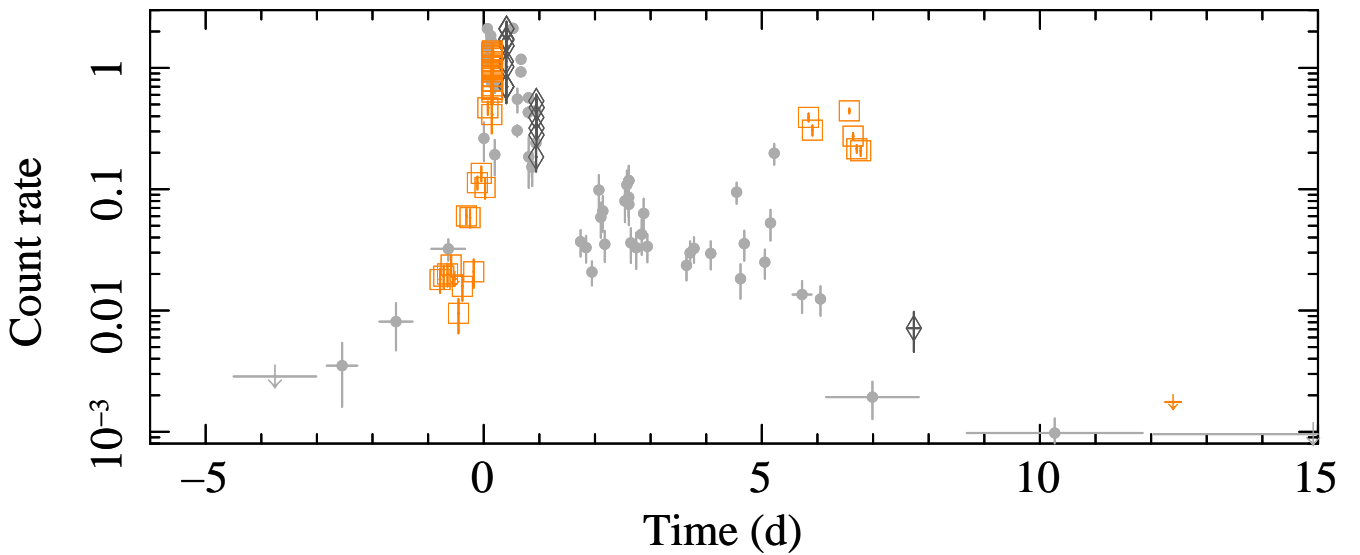}
 		\caption{XRT light curves of the 2007 February 9 ``periastron'' outburst (light grey filled circles), 
	        with superimposed the 2007 July 24 ``apastron'' outburst (dark grey empty diamonds), 
		and the 2008 June 16 ``apastron'' outburst (orange empty squares), folded with a period of 164.6 days.
		The time is relative to the peak of the  2007 February 9 outburst. 
		}
                \label{igr112p3:fig:period_lcv}
	\end{figure*}

\src\ is the  member of the new SFXT class observed more in depth, 
thanks to the known periodicity of its outburst recurrence ($\sim$165 days).
This is an important property among these new X--ray transients, 
which allows to plan a monitoring of the behavior during outbursts.

Here we report on a long observation performed with \sw, which for the first time allows a frequent 
sampling of the X--ray emission during the onset of an outburst from \src. 
This new outburst was expected after 329 days from the last observed one, 
on 2007 July 24,  
and the observations were performed on 2008 June 16,
the day of the expected bright peak.
A total of 15 snapshots were collected. 
After 5--6 days from the bright peak, another short \sw\ observation was performed,
to monitor the declining part of the outburst (bottom panel in Fig.~\ref{igr112p3:fig:lcv}), 
until its end, observed about 12 days after the bright peak, 
when an upper limit to the luminosity could be placed at $\sim$10$^{33}$~erg~s$^{-1}$. 

The source shows a hard X--ray spectrum, well fitted with an absorbed
power-law with a photon index of $\sim$1, consistent with previous outbursts
observed with \sw. There is also no evidence for a variability in the 
absorbing column density during the outburst, and also in comparison
with previously observed outbursts (e.g. Paper~II). 
An epoch folding analysis results in a spin period consistent with previous
determinations, and the folded pulse profiles are similar to those
observed with \xmm\ in 2007 February (Paper~III), with a dependence 
on the source flux. 
Indeed, the pulse profile obtained from XRT observation 001
is very similar to what observed during the bright flare with 
{\it XMM-Newton} (part ``A'' of the observation, Paper~III), 
while the XRT observation 002, which caught the source 
in a state almost an order of magnitude fainter than during 
observation 001, displays a pulse profile very similar to that 
during the fainter state ``B'' with XMM-Newton (Paper~III). 
The different pulse shapes are hence very likely linked to a change 
in flux (almost an order of magnitude).
A change in the pulse profile with the luminosity  
has already been observed in other accreting pulsars \citep{Parmar1989}.

The \src\ historical light curve observed with \sw/XRT is 
reported in Fig.~\ref{igr112p3:fig:historical_lcv}.
The observed maximum of the 2008 outburst is slightly fainter than the 
peaks observed during the two previous outbursts in 2007, but this may be 
due to the fact that the observation ended before the actual peak was reached.

Fig.~\ref{igr112p3:fig:period_lcv} shows the \src\ light curve 
folded on a period of 164.6~days, to properly
compare the properties of the three observed outburst. 
Based on the first recognized periodicity of 329~days (Paper~I)
we have designated 
these three outbursts as ``periastron'' (2007 February) 
and ``apastron'' outbursts (2007 July and
the one reported here).

In Paper~III 
we proposed an alternative explanation for the SFXTs outbursts,
based on the properties of the observed \src\ X--ray light curve.
Its shape could not be modelled by  Bondi-Hoyle accretion from a homogeneous and 
spherically symmetric wind from the blue supergiant companion.
We suggested the presence, besides the symmetric polar wind, of an
equatorially enhanced wind component (denser and slower than the polar one), 
possibly clumpy (to explain the high variability of the source flux) and inclined 
with respect to the orbital plane, to account for the narrow and steep X--ray 
light curve in outburst and the periodic outbursts.
The presence of equatorially enhanced winds in blue supergiants has also been
suggested in literature from the variability of
optical spectra \citep[e.g., ][]{Markova2008}, from simulations \citep{Ud-doula2008}, 
and from X--ray observations of a number of supergiant High Mass X--ray Binaries 
displaying apastron flares
\citep[e.g., ][]{Corbet2006,Pravdo2001,Roberts2001}. 

In Paper~III 
we also discussed two possible orbital geometries for \src, 
and obviously different wind parameters for both the equatorial and the polar components, 
to explain the X--ray luminosity observed:
(1) an orbital period, $P$, of 329~days (and a circular orbit, 
with two observed outbursts per orbit) or 
(2) an orbital period, $P/2$, of $\sim$165~days, with only one outburst
per orbit and a higher eccentricity, together with a possibly  
truncated equatorial wind (this to avoid observing an outburst at $P/4$, 
which is now excluded based on the observation performed in 2008 March, 
see Fig.~\ref{igr112p3:fig:historical_lcv}).

The new 2008 outburst, compared in Fig.~\ref{igr112p3:fig:period_lcv} 
with the one observed in 2007 February and 2007 July, displays a similar 
duration and X--ray luminosity, although the sampling of the three monitoring 
campaigns is very different. 
Thus we will not discuss
further the wind parameters (density and velocity needed to explain the 
X--ray emission), due to the analogies of the three outburst light curves, 
and we will refer to Paper~III, where we report a possible choice for the
wind parameters that match the observed X-ray light curve, 
as the model is essentially open and different combinations of wind density 
and velocity in the equatorial wind can reproduce the X-ray light curve. 
On the other hand, in 2008 the source 
seems to remain longer in a sort of bright state 
 after the peak (see Fig.~\ref{igr112p3:fig:period_lcv}): 
observations performed 5--6 days after the bright peak, 
found  the source still at a high X--ray level ($\sim$0.3\,counts\,s$^{-1}$), 
with a count rate about one order of magnitude higher 
than the average level of the 2007 February declining tail. 
It is important to note, however, that \src\
is a highly variable source, displaying a frequent flaring activity  
with a large dynamic range even during the tail of the 2007 
February outburst (this is clearly shown in Fig.~\ref{igr112p3:fig:period_lcv}). 
Thus we cannot exclude that 
during the last outburst in 2008, \sw\ observations caught by 
chance only the bright peaks of a similar flaring
activity, while the tail,  much fainter on average, was not observed, 
due to the several gaps and sparse sampling.
A second possibility is the variability of the wind properties 
which could produce a longer outburst at periastron 
or the presence of single dense clumps, 
which are not uncommon in blue supergiants \citep[e.g., ][]{Owocki2006} 
 and could have been accreted and thus produced these bright peaks. 
A third yet more unlikely possibility is that this quite high count 
rate observed about 5--6 days after the bright peak on 2008 June 16, 
could be part of a much longer X--ray outburst peak, lasting 5--6 days 
(but note that there is a large observing gap between the two sets of 
observations), due to accretion of matter from the equatorial wind 
near the apastron passage, where the neutron star velocity could be 
5--6 times lower than during the periastron passage (assuming the 
same size for the compressed wind outflow) in an eccentric binary.
This latter hypothesis can be actually excluded by the following 
argument: a supposed range of variability for the neutron star
orbital velocity of 5--6 times can be obtained only in an orbit 
with a high eccentricity, $e\sim$0.7. 
On the other hand, $e=0.7$ would imply a very high dynamic range 
of $\sim60$ (assuming wind parameters discussed in Paper~III)
between the X--ray luminosity  produced at the periastron
passage and that produced at apastron, when the neutron star crosses the 
equatorial wind; but this large difference in the peak amplitude 
(which is similar during all outbursts) is excluded by the observations.

The new observations reported here trace quite well the rise to the new outburst,
and if compared with previous observations, allow to possibly say something more
conclusive about the true orbital period of this binary system.
Indeed we favor an eccentric orbit with an orbital period of $\sim$165~days, with one outburst per orbit
produced near the periastron passage.
This conclusion can be derived from the arguments we discuss below.

The three outburst peaks observed with \sw/XRT can be very well overlaid, when they are 
folded on a period of 164.6~days (see 
Fig.~\ref{igr112p3:fig:period_lcv}). 
This interestingly implies that they are very closely time-locked.

In the framework of our proposed model (where outbursts  
are produced when the neutron star crosses
the supergiant equatorial wind, which is inclined with respect to the orbital plane), 
this close time-locking implies that, if the true
orbital period is 329~days, the orbit should be perfectly circular (to have two outbursts per orbit
at exactly a half of the orbital period). 
Moreover, the supergiant compressed outflow in the magnetic equator  
should be highly stable and ``rigid'' 
up to large distances from the companion: for a binary period of $P=329$~days, the orbital separation 
is $\sim$4$\times$10$^{13}$~cm (about  25 stellar radii for a typical B1 supergiant).  
Detailed studies and simulations of equatorially enhanced supergiant winds suggest
that a rotating supergiant with a dipole magnetic field aligned to the star's rotation
axis, can produce a steady magnetically confined line-driven stellar wind on the equatorial plane
\citep{Ud-doula2008}. 
For example, the wind of a B1-type supergiant can be perturbed to form 
an equatorial density enhancement 
with a relatively low magnetic field of about 20--30\,G. 
Indeed, assuming the ``wind magnetic confinement
parameter'' $\eta$ \citep{Ud-doula2002}, 
the minimal magnetic field, B$_{\rm min}$(for $\eta$=1), for an equatorial wind compression  is
B$_{\rm min} =  \sqrt{{\dot M_{w}} v_\infty}/R_{\rm c}$, where $\dot M_{w}$ is the wind mass loss rate,
$v_\infty$ the terminal velocity and R$_{\rm c}$ the supergiant radius.
Assuming, for example,  $\dot M_{w}$=5$\times$10$^{-7}$~M$_\odot$~yrs$^{-1}$, R$_{\rm c}$=30~R$_\odot$
and $v_\infty$=1000~km~s$^{-1}$, we obtain B$_{\rm min}$=27~G.

\begin{deluxetable*}{llllr}
  \tabletypesize{\scriptsize}
  \tablewidth{0pc} 	      	
  \tablecaption{Observation log.\label{igr112p3:tab:alldata}} 
  \tablehead{
\colhead{Sequence\tablenotemark{a}} & \colhead{Start Time} &  \colhead{Start time (UT)} &  \colhead{End time (UT)} &  
               \colhead{Exposure\tablenotemark{b}} \\
\colhead{}     & \colhead{(MJD)}    & \colhead{(yyyy-mm-dd hh:mm:ss)} & \colhead{(yyyy-mm-dd hh:mm:ss)} &  \colhead{(s)} \\
\colhead{(1)}    & \colhead{(2)}         & \colhead{(3)} & \colhead{(4)}     & \colhead{(5)}        
}
  \startdata
00030881033  &54550.0116 &2008-03-25 00:16:42     &2008-03-25 02:07:57     &1829	\\
00030881034  &54552.0186 &2008-03-27 00:26:47     &2008-03-27 02:18:55     &1941	\\
00090005001  &54633.0057 &2008-06-16 00:08:15     &2008-06-16 23:04:57     &18975	\\
00090005002  &54639.6381 &2008-06-22 15:18:48     &2008-06-23 13:59:17     &6740	\\
00090005003  &54646.0529 &2008-06-29 01:16:08     &2008-06-29 08:01:56     &5410	\\
00030881035  &54651.4232 &2008-07-04 10:09:21     &2008-07-04 11:58:55     &1093	
\enddata 
  \tablenotetext{a}{The previous observations are listed in \citet{Sidoli2008:sfxts_paperI}.}
  \tablenotetext{b}{The exposure time is spread over several snapshots  
	(single continuous pointings at the target) during each observation.}
  \end{deluxetable*}

\begin{deluxetable}{lllll}
  \tabletypesize{\scriptsize}
  \tablewidth{0pc} 	      	
  \tablecaption{Spectral fit results.\label{igr112p3:tab:specfits}} 
  \tablehead{
\colhead{Spectrum} & \colhead{$N_{\rm H}$} & \colhead{$\Gamma$} & \colhead{$\chi^2_{\nu}$ (d.o.f.)/} 
                                                         & \colhead{$L_{1 - 10 {\rm \,keV}}$\tablenotemark{a}} \\
\colhead{}         & \colhead{(10$^{22}$ cm$^{-2}$)} & \colhead{} &  \colhead{C-stat(\%)\tablenotemark{b}}  &
                                            \colhead{erg s$^{-1}$)}   \\
\colhead{(1)}    & \colhead{(2)}         & \colhead{(3)} & \colhead{(4)}    & \colhead{(5)}      
}
  \startdata
001 (orb 1--13) & $1.32_{-0.36}^{+0.47}$       & $1.30_{-0.29}^{+0.33}$ & $0.788/18$          & 0.129  \\ %0.1292 10^35
001 (orb 14)     & $3.13_{-1.76}^{+2.42}$       & $1.24_{-0.80}^{+0.89}$ & 384.2   (82.0\%)   & 2.517  \\ %0.2517 10^36
001 (orb 15)     & $1.77_{-0.44}^{+0.55}$       & $0.90_{-0.25}^{+0.27}$ & $0.861/40$         & 5.003  \\ %0.5003 10^36
002 (orb 1)      & $1.96_{-0.66}^{+0.80}$       & $1.58_{-0.44}^{+0.47}$ & 580.5   (82.3\%)   & 1.469  \\ %0.1469 10^36
002 (orb 2)      & $1.29_{-0.60}^{+0.79}$       & $1.05_{-0.47}^{+0.52}$ & 475.5   (65.7\%)   & 1.339  \\ %0.1339 10^36 ergs/s
002 (orb 3)      & $0.88_{-0.25}^{+0.31}$       & $0.85_{-0.22}^{+0.24}$ & $0.904/26$         & 2.031  \\ %0.2031 10^36
002 (orb 4)      & $0.69_{-0.24}^{+0.29}$       & $0.95_{-0.26}^{+0.27}$ & 713.4   (65.0\%)   & 1.230  \\ %0.1230 10^36
002 (orb 5)      & $0.91_{-0.32}^{+0.51}$       & $1.21_{-0.37}^{+0.40}$ & 607.0   (50.0\%)   & 0.822 \\ %0.8220 10^35
002 (orb 6)      & $0.80_{-0.41}^{+0.58}$       & $0.63_{-0.38}^{+0.42}$ & 548.3   (62.0\%)   & 1.270 \\ %0.1270 10^36
002 (all)        & $1.05_{-0.18}^{+0.21}$       & $1.06_{-0.14}^{+0.14}$ & $0.946/81$         & 1.324 \\ %0.1324 10^36
\enddata 
  \tablenotetext{a}{Luminosity in the 1--10\,keV band in units of $10^{35}$ erg s$^{-1}$ obtained 
from the spectral fits adopting a distance of 6.2\,kpc. }
  \tablenotetext{b}{Cash statistics (C-stat) and percentage of Monte Carlo realizations 
that had statistic $<$ C-stat. We performed $10^4$ simulations.}
  \end{deluxetable}

The outer edge of this rigid structure largely depends on the surface magnetic field
of the supergiant, and can reach a few stellar radii. It extends to 20 stellar radii only
in case of highly magnetic Bp stars \citep{Ud-doula2003}. 
Thus, such a large compressed equatorial wind structure seems to be less
likely for a B supergiant. 
This would favor a less wide and eccentric 
orbit for \src\ with an orbital period of $\sim$165 days. 
This also avoids the intersection of the equatorial wind component near apastron passage.

Another piece of evidence favoring the 165 days orbital period
 comes from the fact that there is no evidence for
a changing absorption column density between the three different outbursts
observed with \sw.
If the outbursts  
are produced by sudden accretion of denser material due to the neutron star ingress
into the supergiant equatorial wind, a different absorbing column density should 
in principle be observed between two consecutive outbursts (one at the periastron 
and the other one at the apastron, if we assume $P=329$~days). Since the 
supergiant equatorial component is inclined with respect to the neutron star orbit,
the observer should see one of the two outbursts in front of it, and the other
one from behind it (thus with a higher column of absorbing matter towards the line of sight). 
But a difference in the column density is not observed, within the uncertainties. 
This difficulty could be avoided in the $P=329$ days scenario only if we assume 
that the system is viewed face-on, 
and with a supergiant equatorial wind component perpendicular to the orbital plane. 

Moreover, it seems quite unlikely that the supernova explosion which produced the neutron 
star has completely misaligned the orbital plane with the supergiant spin axis, but did not produce 
at the same time an eccentric orbit, thus leaving behind a perfectly circular orbit (as required if the
true period is 329~days).

In conclusion, all these arguments based on the new  \sw/XRT observations indicate that
an orbital period of 329~days would imply a too much fine-tuned geometry of the system,
and on the other hand suggest that
\src\ is very likely in an eccentric orbit with an orbital period of 164.6~days, where the outbursts
are produced near the periastron passage by the same wind structure crossed every time.

The rise to the outburst is much better sampled in 2008
than during the February 2007 outburst, thus allowing us 
to confirm once more that the shape of the light curve cannot be explained by accretion 
from a spherical distribution of 
matter (homogeneous or not) as already demonstrated in Paper~III.
In particular, the new light curve clearly shows a rise to the outburst of $\sim$2 orders of magnitude  
in the source count rate in less than about $\Delta \phi = 0.002$ orbital phase (assuming $P=164.6$~days period).
On the contrary,
a wind with a spherical distribution (and the same stellar parameters as in Paper~III)
can at most produce this difference in count rate during  $\Delta \phi \sim 0.02$ orbital phase, assuming 
the highest possible orbital eccentricity of e=0.89 (or the neutron star orbit would lie
inside the supergiant at periastron).
\citet{Negueruela2008} proposed different possible geometries for SFXTs, within the framework
of a clumpy spherical wind from the supergiant companion.
In their model, a periodic recurrence of the outbursts could be explained only with an
eccentric orbit. On the other hand, even assuming a very high eccentricity of the orbit, 
as discussed above, any spherically symmetric distribution of matter
(although clumpy) cannot reproduce both the steep rise to the outburst and the entire narrow shape of
the outburst in \src\ (Paper~III). 
Our interpretation of the outbursts of \src\ with a preferential plane for
the wind outflow can benefit from much better constrained wind parameters (density, velocity,
mass loss rate), and its application to all the other SFXTs needs to be tested and confirmed.
Indeed, we are currently monitoring 4 SFXTs (see, e.g., \citealt{Sidoli2008:sfxts_paperI}) 
with frequent observations spread over a long-baseline in order to assess whether 
it is possible to extend or reject this model for all the other members of the same class.

%%%%%%%%%%%%%%%%%%%%%%%%%%%%%%%%%%%%%%%%%%%%%
\acknowledgements
We thank the \sw\ team for making these observations possible,
in particular Scott Barthelmy (for his invaluable help with BAT), 
the duty scientists,  and science planners. 
PR thanks Valentina La Parola for very helpful discussions. 
LS thanks INAF-IASF Palermo, where most of the work was carried out, 
for their kind hospitality. 
LS thanks Asif ud-Doula and Stan Owocki for very helpful discussions.
H.A.K.\ was supported by the \sw\ project. 
This work was supported by contract ASI/INAF I/088/06/0 and I/023/05/0.

%%%%%%%%%%%%%%%%%%%%%%%%%%%%%%%%%%%%%%%%%%%

\end{document}